\documentclass[showpacs,amsmath,amssymb, aps, prd, superscriptaddress, nofootinbib, showkeys, 11pt]{revtex4-1}

\usepackage{amsfonts}
\usepackage{hyperref}
\usepackage{graphicx}
\usepackage{keyval}
\usepackage{caption}
\usepackage{subfig}

\newcommand{\dint}{\int}

\begin{document}

\author{Bertrand Chauvineau}\email{Bertrand.Chauvineau@oca.eu}\affiliation{Laboratoire Lagrange (UMR 7293), Universit\'{e} de Nice-Sophia Antipolis, CNRS, Observatoire de la C\^{o}te d'Azur, BP 4229, 06304 Nice cedex 4, France}\author{Davi C. Rodrigues}\email{davi.rodrigues@cosmo-ufes.org}\affiliation{Departamento de F\'{\i}sica, Universidade Federal do Esp\'{\i}rito Santo, Av. Fernando Ferrari 514, Vit\'{o}ria, ES, 29075-910 Brazil} \author{J\'{u}lio C. Fabris}\email{julio.fabris@cosmo-ufes.org}\affiliation{Departamento de F\'{\i}sica, Universidade Federal do Esp\'{\i}rito Santo, Av. Fernando Ferrari 514, Vit\'{o}ria, ES, 29075-910 Brazil}
\title{Scalar-tensor theories with an external scalar}

\begin{abstract}
Scalar-tensor (ST) gravity is considered in the case where the scalar is an
external field. We show that General Relativity (GR) and usual ST gravity
are particular cases of the External Scalar-Tensor (EST) gravity. It is
shown with a particular cosmological example that it is possible to join a
part of a GR solution to a part of a ST one such that the complete solution
neither belongs to GR nor to ST, but fully satisfies the EST field
equations. We argue that external fields may effectively work as a type of
screening mechanism for ST theories.
\end{abstract}

\pacs{04.50.Kd, 98.80.-k} 

\keywords{Scalar-tensor gravity, external fields, Robertson-Walker cosmology, screening mechanism}

\maketitle

\tableofcontents

\section{Introduction}

Scalar-Tensor (ST) gravity theories constitute one of the main sources of
extensions to General Relativity (GR) (for reviews see \cite{w14,fm03,f04,cf11}). They constitute a class of theories with well posed
actions where the gravitational fields are given by both the metric and a
scalar field. Diverse physical motivations for GR extensions naturally lead
to ST actions \cite{fm03, f04, cf11}. On the other hand, currently there is no
solid observational or experimental evidence that clearly favours a ST
model over GR. This has lead to considerations on mechanisms that could
somehow hide the non-observed scalar field effects in the Solar system,
while allowing it to develop a nontrivial role at different scales, say at
the cosmological level. Such mechanisms, which change the effective number
of degrees of freedom of gravity, by hiding the scalar field at appropriate
scales, have received considerable attention recently
\cite{v72, kw04, hk10, kmz12}.

External fields appear in diverse contexts in physics. In particular, an
external field is sometimes defined as a field whose expression as a spacetime
function is known beforehand. Thence the role of the field equations is to
provide solutions to the other (non-external) fields. From the action
perspective, an external field is a quantity that should not be varied,
since it is already known. For instance, the Minkowski metric in the context
of electromagnetism in flat spacetime is an external rank two field. In this
case the spacetime geometry is fixed beforehand and one uses the field
equations, which are derived from variations with respect to other fields,
in order to derive electromagnetic configurations that are compatible with
such geometry.

It is not a novelty that external fields are used as part of the framework of gravity theories. For instance, in \cite{rw04, sss05, rls10} the authors, based on Renormalization Group expectations, argue in favor of a gravitational action with two external scalar fields ($G$ and $\Lambda$). This is a natural starting point since the dynamics of these two fields, in this context, should not be inferred from pure classical field equations, the Renormalization Group equations together with certain ansatz are used to partially fix the dynamics of $G$ and $\Lambda$ (for recent developments see \cite{rcp15}). Another proposal that uses external fields is that of  massive gravity in arbitrary backgrounds. The latter is  equivalent to bigravity in the limit in which the coupling constant of one of the rank-2 tensors goes to infinity (see e.g. \cite{Hassan:2011tf}, see also \cite{Isham:1971gm, Damour:2002wu, deRham:2010kj, Hassan:2011vm}). These actions contain a usual rank-2 tensor together with an additional one that has external character, that is the full field equations do not consider an equation coming from the action variation with respect to the external tensor (which may or may not describe a Minkowski spacetime geometry). We also cite the case of unimodular gravity (see \cite{Weinberg:1988cp, Unruh:1988in, Henneaux:1989zc, Ellis:2010uc, Kluson:2014esa, FernandezCristobal:2014jca} and references therein). In unimodular gravity, when formulated as in \cite{Weinberg:1988cp}, there is no field that is fully external, but the determinant of the metric behaves as an external field. That is, the metric has the same role as in the Einstein-Hilbert action, except that its determinant has a fixed value.\footnote{In the context of unimodular gravity, it is common to say that the metric determinant is ``non-dynamical'' (e.g., \cite{Weinberg:1988cp}), but the label of ``external'', as used here, also applies  to it and it has been used in some references, in particular see \cite{Henneaux:1989zc}.} Its action can also be recast by using a Lagrange multiplier such that the metric is fully non-external, but a condition is imposed between the metric determinant and an {\it external} scalar density field \cite{Henneaux:1989zc, FernandezCristobal:2014jca}. As a last example, the more recently proposed approach to modify gravity inspired on continuous mechanics \cite{bt13}, is closely related to the approach developed here. In Ref. \cite{bt13} an external tensor (which is not a metric) is also used. Contrary to Ref. \cite{bt13}, our emphasis here will be on an external scalar and the comparison between the frameworks of GR and ST gravity to EST gravity, that is, a ST gravity whose scalar is an external field.

In order to be clear, the definition that is here used for an external field is the following: it is a field whose variation inside the action should not be considered in the process of deriving the field equations. Hence, for a fixed number of fields, the higher is the number of external fields, the lower is the number of field equations derived from the action.  The main concern in this work is precisely to look for EST solutions that are different from both GR and ST solutions. It turns out that EST theories allow a different way of recovering GR behaviour in some spacetime regions. The transition between GR to non-GR behaviour depends on the boundary conditions associated with the system.

This work is organised as follows: Section \ref{sec2} is devoted both to the introduction of EST gravity and to address general differences and similarities to ST gravity and GR. A particular cosmological solution that exhibits both a GR phase and a ST one (but that is globally neither a GR nor a ST solution) is proposed in Section \ref{sec3}. In Section \ref{sec4} we present our conclusions and in Appendices \ref{appA}, \ref{appB}, \ref{appC} we present respectively issues on the stress-energy conservation related with external fields, additional details on the proposed particular solution and an extension of the latter for barotropic fluids. A mechanical example on the use of external variables is presented in Appendix \ref{appMec}, and Appendix \ref{appE} shows a first test using SN data.

Commonly used abbreviations: BD = Brans-Dicke, EST = Scalar-Tensor with an external scalar field, GR = General Relativity,  RW = Robertson-Walker, ST = Scalar-Tensor.

\bigskip

\section{Scalar-tensor gravity with an external scalar field} \label{sec2}

Consider the following action in which $\Phi $ is an external scalar field%
\begin{eqnarray}
S\left[ g,\Psi \right] &=&\frac{1}{16\pi }\dint \left( \Phi R-\frac{\omega
\left( \Phi \right) }{\Phi }g^{ab}\partial_{a}\Phi \partial_{b}\Phi - 2\Phi \Lambda \left( \Phi \right) \right) \sqrt{-g}d^{4}x+S_{m}\left[ g,\Psi %
\right],  \label{ST action}
\end{eqnarray}%
where $\Psi $ collectively represents matter fields. It looks as the usual
ST action, but the difference (at the action level) is that the scalar field 
$\Phi $ is not to be varied, since it is external (as defined in the
Introduction). For this reason, the action dependence in $\Phi $ does not
explicitly enter the notation $S\left[ g,\Psi \right] $. Thence, apart from
the matter $\Psi $ field equations, the only equations derived from the
action are given by%
\begin{equation}
\Phi \left( R_{ab}-\frac{1}{2}Rg_{ab}+\Lambda g_{ab}\right) =8\pi T_{ab}+%
\frac{\omega }{\Phi }\left( \partial _{a}\Phi \partial _{b}\Phi -\frac{1}{2}%
g_{ab}g^{cd}\partial _{c}\Phi \partial _{d}\Phi \right) +\nabla _{a}\partial
_{b}\Phi -g_{ab}\square \Phi .  \label{Einstein ST}
\end{equation}%
Since (i) $\Phi $\ does not appear in the matter sector of the action (\ref%
{ST action}), (ii) all the $\Psi $-dependent terms are in $S_{m}$, and (iii)
none of the fields entering $S_{m}$\ is external, one gets the stress tensor
conservation%
\begin{equation}
\nabla ^{a}T_{ab}=0  \label{stress conservation}
\end{equation}%
as a direct consequence of the diffeomorphism invariance of $S_{m}$\ and of
the fact that each field $\Psi $\ generates a Lagrange equation (see
Appendix \ref{appA} for the requirement of each assumption).

 Since $\Phi$ is external, there is no field equation derived from the action variation with respect to $\Phi$.\footnote{Similarly, for electromagnetism within an {\it external} spacetime geometry (that can be Minkowski in particular), one can use the action $S[A]=-\frac 14 \int F_{a b} F^{a b} \sqrt{- g} \, d^4x$ to derive the field equations. These are derived  by setting $\delta S / \delta A^a =0$. Note that the equation $\delta S / \delta g^{a b} =0$ is wrong in general in this context.} Thence, equations (\ref{Einstein ST}) and (\ref{stress conservation}) are the only equations of the theory. Thanks to (\ref{stress conservation}), the contracted Bianchi identity and the covariant derivatives commutation rule, the divergence of (\ref{Einstein ST}) leads to%
\begin{equation}
\left[ 2\frac{\omega }{\Phi }\square \Phi +\partial _{c}\Phi \partial
^{c}\Phi \frac{d}{d\Phi }\left( \frac{\omega }{\Phi }\right) +R-2\frac{d}{%
d\Phi }\left( \Phi \Lambda \right) \right] \partial _{b}\Phi =0.
\label{EST induced scalar eq}
\end{equation}%
Contracting (\ref{Einstein ST}) and eliminating $R$, one gets%
\begin{equation}
\left[ \square \Phi -\frac{8\pi T-\omega ^{\prime }\partial _{c}\Phi
\partial ^{c}\Phi -2\Phi \left( \Lambda -\Phi \Lambda ^{\prime }\right) }{%
2\omega +3}\right] \partial _{b}\Phi =0.  \label{ext scalar field eq}
\end{equation}%
We remark that this equation is \textit{not} the scalar field equation that is valid in usual ST gravity,\footnote{This is obtained by varying (\ref{ST action}) with respect to $\Phi$, where $\Phi$ is \textit{not} an external field in the usual ST framework.} because of the presence of $\partial \Phi $ as an overall factor. In the case that $\partial \Phi \not= 0$ everywhere, then eq. (\ref{ext scalar field eq}) leads to the usual ST equation
\begin{equation}
\square \Phi =\frac{8\pi T-\omega ^{\prime }\partial _{c}\Phi \partial
^{c}\Phi -2\Phi \left( \Lambda -\Phi \Lambda ^{\prime }\right) }{2\omega +3}.
\label{usual scalar eq}
\end{equation}%
The above equation is precisely the same equation that is derived from the action (\ref{ST action}) variation with respect to $\Phi$. It then should be clear that the equation set (\ref{Einstein ST}), (\ref{stress conservation})  and (\ref{usual scalar eq}) composes the usual ST fundamental field equations \cite{w14, w93}, while the equation set (\ref{Einstein ST}) and (\ref{stress conservation})  composes the EST fundamental field equations (from which (\ref{ext scalar field eq}), that is less constraining than (\ref{usual scalar eq}), is just a logical consequence). Let us also stress that, in the usual ST case, the contracted Bianchi identity just results in a triviality, since equation (\ref{usual scalar eq}) is already known as a fundamental field equation. On the other hand, in the EST case, the contracted Bianchi identity leads to equation (\ref{ext scalar field eq}), that is worth to be considered, since it does not appear explicitly (while it is implicitly present) in the set of fundamental field equations. Therefore, the EST action (\ref{ST action}), together with the assumption that $\partial \Phi \neq 0$ everywhere, is equivalent to a ST theory, since in this case, the field equations solutions are the same of the usual ST (provided that the boundary conditions and the functions $\omega \left( \Phi \right) $ and $\Lambda \left( \Phi \right) $ are the same). Thence, the solutions of ST theory are particular solutions of the corresponding EST theory. However, (\ref{ext scalar field eq}) differs from the usual scalar equation by the scalar field gradient present as an overall factor. As a consequence, independently on the $\omega \left( \Phi \right) $ and $\Lambda \left( \Phi \right) $ functions, (\ref{ext scalar field eq}) also admits solutions such that
\begin{equation}
\partial _{b}\Phi =0.  \label{GR condition}
\end{equation}%
Reinserting in (\ref{Einstein ST}), it immediately results that the EST field equation (\ref{Einstein ST}) becomes identical to GR field equation, with the Newton's constant $G$ numerically given by $1/\Phi$. Thence, GR solutions also generate EST solutions with the same stress tensor, the EST metric being the GR's one and the scalar field being $\Phi =1/G$, where $ G$\ is Newton's constant. 

To be complete, it could be worth identifying the cases where EST gravity reduces to the usual ST case. The difference between the two theories just results from the presence of $\partial \Phi $ as an overall factor in (\ref{ext scalar field eq}). Thence, the reduction to usual ST occurs if, and only if, the solutions of $\partial \Phi =0$ also solve eq. (\ref{usual scalar eq}). This conditions reads (considering finite $\omega $ and $\omega ^{\prime }$)
\begin{equation}
\Phi ^{3}\frac{d}{d\Phi }\left( \frac{\Lambda }{\Phi }\right) +4\pi T=0.
\end{equation}%
It is interesting to point out that, if no cosmological term enters the
theory (i.e., if $\Lambda =0$), then EST and ST are equivalent if $T=0$. Hence, in particular, vacuum solutions of EST gravity with $\Lambda(\Phi)=0$ are always ST gravity solutions. For instance, in this case, the propagation of gravitational waves in vacuum obeys the same equations in both theories. 

\bigskip

Another issue to be stressed is (and this is the main point of this paper) is that the set of EST solutions is larger than the set of the corresponding ST or GR solutions. Indeed, one can also derive EST solutions by matching parts of usual ST solutions to parts of GR solutions, if the relevant conditions are satisfied. These conditions ensure  that the EST field equations are valid at all the spacetime points (this will be exemplified in the next section).  Reciprocally, it results from eq. (\ref{ext scalar field eq}) that the general EST solution can only be built from joined ST and GR solutions. Let us remark that, in an EST solution that contains two disjoint spacetime regions where $\partial \Phi =0$, the solutions in these regions satisfy the GR equation, but  in general with different numerical values for the Newtonian constant.

\section{A cosmological solution} \label{sec3}

Let us consider in this section a specific EST action characterized by $%
\Lambda =0$ and a constant $\omega $. The usual ST version of this case is
the original BD theory \cite{bd61}. Consider a flat dust filled RW Universe of
metric%
\begin{equation}
ds^{2}=-dt^{2}+a\left( t\right) ^{2}\delta _{ij}dx^{i}dx^{j}.
\label{flat RW metric}
\end{equation}%
The EST equations to be satisfied by the scale factor $a\left( t\right) $,
the scalar field $\Phi \left( t\right) $ and the dust density $\epsilon
\left( t\right) $ are (\ref{Einstein ST}) and (\ref{stress conservation}).
The non-trivial Einstein's equation components can be written as%
\begin{eqnarray}
3\Phi \frac{a^{\prime 2}}{a^{2}} &=&8\pi \epsilon +\frac{\omega }{2}\frac{%
\Phi ^{\prime 2}}{\Phi }-3\frac{a^{\prime }}{a}\Phi ^{\prime },
\label{EST cosmo eq1} \\[.1in]
2\frac{a^{\prime \prime }}{a}+\frac{\Phi ^{\prime \prime }}{\Phi } &=&-\frac{%
\omega }{2}\frac{\Phi ^{\prime 2}}{\Phi ^{2}}-2\frac{a^{\prime }}{a}\frac{%
\Phi ^{\prime }}{\Phi }-\frac{a^{\prime 2}}{a^{2}}.  \notag
\end{eqnarray}%
Where the prime represents derivation with respect to time. The stress tensor
conservation yields%
\begin{equation}
\left( \epsilon a^{3}\right) ^{\prime }=0.  \label{EST cosmo eq2}
\end{equation}%
For the present case, eq. (\ref{ext scalar field eq}) reads
\begin{equation}
\left[ \left( a^{3}\Phi ^{\prime }\right) ^{\prime }-\frac{8\pi \epsilon
a^{3}}{2\omega +3}\right] \Phi ^{\prime }=0.  \label{EST cosmo eq3}
\end{equation}%
The above equation is implicitly contained in the system (\ref{EST cosmo eq1}) and (\ref{EST cosmo eq2}).  We stress that, within the usual BD framework, the scalar field equation (\ref{usual scalar eq}) should be considered, and this equation is not compatible with $\Phi'=0$. Both the GR and BD solutions of the flat dust filled RW Universe are known, and are solutions of (\ref{EST cosmo eq1}) and (\ref{EST cosmo eq2}), as it can easily be checked. After reminding these solutions in the first subsection and defining the useful related notations, we built in the second subsection a non-trivial EST RW solution got joining a part of the GR solution to a part of the BD one.

\subsection{General Relativity and Brans-Dicke like solutions}\label{sec3.1}

The GR solution reads $a_{GR}\left( t\right) =A\left( t-t_{0}\right)^{2/3}$, where $A$ and $t_{0}$ are integration constants, $t_{0}$\ being the GR big bang time. From the EST perspective, this solution can be achieved if the field $\Phi$ is set to be a constant. Thence, we write the corresponding EST solution
in the form%
\begin{eqnarray}
a_{GR}\left( t\right) &=&A\left( t-t_{0}\right) ^{2/3},  \label{GR phase} \\[.1in]
\Phi _{GR}\left( t\right) &=&C,  \notag
\end{eqnarray}%
where $C$\ is a constant.

The BD solution was found in \cite{gfr73}. It can be equivalently rewritten as%
\begin{eqnarray}
a_{BD}\left( t\right) &=&B\left( t-t_{+}\right) ^{q_{+}}\left(
t-t_{-}\right) ^{q_{-}},  \label{BD phase} \\[.1in]
\Phi _{BD}\left( t\right) &=&\Phi _{0}\left( t-t_{+}\right) ^{p_{+}}\left(
t-t_{-}\right) ^{p_{-}}  \notag
\end{eqnarray}%
with%
\begin{eqnarray}
q_{\pm } &=&\frac{1+\omega \mp s\sqrt{1+\frac{2}{3}\omega }}{4+3\omega }
\label{q & p}, \\[.1in]
p_{\pm } &=&\frac{1\pm 3s\sqrt{1+\frac{2}{3}\omega }}{4+3\omega }.  \notag
\end{eqnarray}%
In the above, $B$, $\Phi _{0}$, $t_{-}$ and $t_{+}$ are (independent) integration
constants, and $s$ is a sign. Remark that we can decide that $t_{+}>t_{-}$
without loss of generality (permuting $t_{+}$\ and $t_{-}$\ is equivalent to
change $s$ in $-s$). We make this choice in the following, with the
consequence that $t_{+}$\ is the BD singular time for observers living in
the $t>t_{+}$\ phase\footnote{%
For observers living in the phase $t>t_{+}$, this singular time corresponds
to a big bang time whatever $s$ in the case $\omega >0$, but only if $s=-1$
in the case $\omega \leq 0$.}. Let us also remark that%
\begin{equation}
a_{BD}^{3}\Phi _{BD}^{\prime }=\Phi _{0}B^{3}\left[ \frac{2}{4+3\omega }%
t-\left( p_{+}t_{-}+p_{-}t_{+}\right) \right]  \label{scal eq1}
\end{equation}%
while the scalar equation (\ref{usual scalar eq}) reads%
\begin{equation}
\left( a^{3}\Phi _{BD}^{\prime }\right) ^{\prime }=\frac{8\pi \epsilon a^{3}%
}{3+2\omega }.  \label{scal eq2}
\end{equation}%
These two equations yield, reminding that $\epsilon a^{3}$ is constant
thanks to the energy conservation%
\begin{equation}
\frac{8\pi \epsilon a^{3}}{3+2\omega }=\frac{2\Phi _{0}B^{3}}{4+3\omega }.
\label{from sc eq1&2}
\end{equation}%
For simplicity, let us just consider scenarios with $\epsilon \geq 0$ (that
is always the case considering ordinary matter) and $\Phi _{0}\geq 0$
(meaning that the local effective gravitational constant $G_{eff}=\frac{%
4+2\omega }{3+2\omega }\frac{1}{\Phi _{BD}}$\ is positive). The solutions (%
\ref{BD phase}) that fulfil these requirements exist if and only if $\omega $%
\ satifies%
\begin{equation}
4+3\omega >0.  \label{omega condition}
\end{equation}

Let us point out that the limit for $\omega \longrightarrow \infty $ of the
BD solution (\ref{BD phase}) reads, if $t_{+}$\ and $t_{-}$\ are considered
constant in the limit process (which makes sense since there is no
constraint on $t_{+}$\ and $t_{-}$\ for (\ref{BD phase}) to be a BD solution),
\begin{eqnarray}
a_{BD\lim }\left( t\right) &=&B\left( t-t_{+}\right) ^{1/3}\left(
t-t_{-}\right) ^{1/3},  \label{BD phase lim} \\[.1in]
\Phi _{BD\lim }\left( t\right) &=&\Phi _{0}.  \notag
\end{eqnarray}%
This does not correspond to (\ref{GR phase}) unless requiring $t_{+}=t_{-}$.
As pointed in \cite{c07} this is due to the fact that, generically, the limit of
BD dust cosmology for $\omega \longrightarrow \infty $ is not just dust
filled GR cosmology, but GR cosmology with a matter sector containing a
massless scalar field besides the original dust field. The presence of such
a residual massless scalar is generic in such a limit process: it results
from the fact that the BD action term $\omega \Phi ^{-1}\left( \partial \Phi
\right) ^{2}$ has no trivial limit when $\omega \longrightarrow \infty $,
the scalar gradient $\partial \Phi $\ going then to zero \cite{c03, c07}. Accordingly, the residual scalar field entering the solution (\ref{BD phase lim}) is zero in the case $t_{+}=t_{-}$, and in this case only.\footnote{This point is not just formal, it is physically meaningful. For instance, the age of a  dust filled flat RW Universe within GR reads  $T_{GR}=2/\left( 3H_{0}\right) $. On the other hand, the age of a dust filled flat RW Universe within BD, resulting from (\ref{BD phase}) and $\omega \longrightarrow \infty $   differs from GR, since all the values in the interval $]1/(3H_0),2/(3H_0)]$ can be achieved \cite{c07}.}

\subsection{A nontrivial External Scalar-Tensor solution}\label{sec3.2}

We develop here a nontrivial cosmological EST solution with $\omega > -4/3$. This solution is composed by two phases, one which is part of a GR solution (at earlier times, eq. (\ref{GR phase})), and the other a ST one (at late times, eq. (\ref{BD phase})). The  instant that joins both solutions is labelled $t_m$. It should be stressed  that in order for this junction to be an EST solution, eqs. (\ref{EST cosmo eq1}) and (\ref{EST cosmo eq2}) need to be satisfied at any instant, including the instant $t_m$. The whole problem then depends on the eight parameters $\left( A,C,t_{0},t_{m},B,\Phi _{0},t_{+},t_{-}\right) $. Since both $a^{\prime \prime }$ and $\Phi^{\prime \prime }$\ explicitly enter the system  (\ref{EST cosmo eq1}) and (\ref{EST cosmo eq2}), and since the matter distribution is regular, the continuity of the scale factor, the scalar field and their first derivatives is required. This constrains four of these quantities to be dependent of the four other ones. The scenario being a BD-like expansion following an earlier GR-expansion, it is natural to express the four BD parameters $\left( B,\Phi_{0},t_{+},t_{-}\right) $ as a function of the three GR parameters $\left( A,C,t_{0}\right) $ and the matching time $t_{m}$, that is left arbitrary. Remark that the invariance of the problem with respect to time translation allows to chose arbitrarily one of the instants entering the problem. It is natural to adopt the global solution singularity as the origin of time, i.e. to set $t_{0}=0$. The matching time then satisfies $t_{m}>0$. It should also satisfy $t_{m}>t_{+}$, otherwise there is no matching and no place for the GR phase. 

Before proceeding, we remark that the proposed solution union may sound as a matching problem as discussed in \cite{mtw73, bv81}, in particular the problem of matching a Schwarszchild spacetime region with an RW one
\cite{nh84, h85}, that results in the so called Einstein-Straus vacuole solution
\cite{b00}. (For matching conditions in ST theories, see for instance \cite{nkty14}.)
It is worth pointing out the differences to the current problem. The metric
and coordinates as written in (\ref{flat RW metric}) represent the solution
in the whole spacetime, and every solution of the system (\ref{EST cosmo eq1}%
) and (\ref{EST cosmo eq2}) is an EST solution in the whole spacetime. It is necessary and sufficient that the equations (\ref{EST cosmo eq1}) and (\ref{EST cosmo eq2}) are satisfied, in particular this implies that one should just verify if these equations are satisfied for $t<t_{m}$, $t>t_{m}$ and at $t=t_{m}$. The proposed solution satisfies this criteria.
Within the present context, for $t<t_m$ the EST solution comes from eq. (\ref{GR phase}), and for $t>t_m$ from eq. (\ref{BD phase}). Hence, the related dust densities are given by
\begin{eqnarray}
6\pi \epsilon _{GR}\left( t\right)  &=&\frac{C}{t^{2}},
\label{dust densities} \\[.1in]
4\pi \epsilon _{BD}\left( t\right)  &=&\frac{3+2\omega }{4+3\omega }\frac{%
\Phi _{0}}{\left( t-t_{+}\right) ^{3q_{+}}\left( t-t_{-}\right) ^{3q_{-}}}. 
\notag
\end{eqnarray}%
The continuities of the scalar and its (logarithmic) derivative, of the
scale factor and its (logarithmic) derivative, give respectively%
\begin{equation}
\Phi _{0}\left( t_{m}-t_{+}\right) ^{p_{+}}\left( t_{m}-t_{-}\right)
^{p_{-}}=C,  \label{scalar cont}
\end{equation}%

\begin{equation}
\frac{p_{+}}{t_{m}-t_{+}}+\frac{p_{-}}{t_{m}-t_{-}}=0,
\label{scalar-der cont}
\end{equation}%

\begin{equation}
B\left( t_{m}-t_{+}\right) ^{q_{+}}\left( t_{m}-t_{-}\right)
^{q_{-}}=At_{m}^{2/3},  \label{scalefact cont}
\end{equation}%

\begin{equation}
\frac{q_{+}}{t_{m}-t_{+}}+\frac{q_{-}}{t_{m}-t_{-}}=\frac{2}{3t_{m}}.
\label{scalefact der cont}
\end{equation}%
Since $p_{+}+p_{-}=2/\left( 4+3\omega \right) $, one gets from (\ref%
{scalar-der cont})%
\begin{equation}
t_{m}-t_{+}=-\frac{4+3\omega }{2}p_{+}\left( t_{+}-t_{-}\right) .
\label{tm minus tplus}
\end{equation}%
Since $t_{m}>t_{+}$ and $4+3\omega >0$, this condition implies that $p_{+}<0$
since we imposed $t_{+}>t_{-}$. Thence the sign $s$ is necessarily $-1$, and
the $q_{\pm }$\ and $p_{\pm }$ exponents read%
\begin{eqnarray}
q_{\pm } &=&\frac{1+\omega \pm \sqrt{1+\frac{2}{3}\omega }}{4+3\omega },
\label{q & p with s=-1} \\[.1in]
p_{\pm } &=&\frac{1\mp 3\sqrt{1+\frac{2}{3}\omega }}{4+3\omega }  \notag
\end{eqnarray}%
and have the following signs: $q_{-}>0$ if and only if $\omega >0$,\ $q_{+}>0
$, $p_{-}>0$ and\ $p_{+}<0$.

The quantities $t_{+}$ and $t_{-}$\ can be derived from  (\ref{scalar-der cont}), (\ref{scalefact der cont}) and (\ref{q & p with s=-1}). One gets%
\begin{equation}
t_{\pm }=\frac{1}{1\pm 3\sqrt{1+\frac{2}{3}\omega }}t_{m}.
\label{BD param tpm}
\end{equation}%
From the above, and since $\omega >-4/3$, one sees that $t_{-}<0$. One also
sees from (\ref{BD param tpm}) that $t_{+}<t_{m}$, meaning that the matching
really occurs without any further condition on $t_{m}$. The quantities $B$
and $\Phi _{0}$\ follow from (\ref{scalar cont}) and (\ref{scalefact cont}).
One gets%
\begin{widetext}
\begin{eqnarray}
B &=&A\left( 3\sqrt{1+\frac{2}{3}\omega }+1\right) ^{q_{+}}\left( 3\sqrt{1+%
\frac{2}{3}\omega }-1\right) ^{q_{-}}\left( 9+6\omega \right) ^{-\frac{%
1+\omega }{4+3\omega }}t_{m}^{\frac{2}{3\left( 4+3\omega \right) }},
\label{BD param Bphi} \\[.1in]
\Phi _{0} &=&C\left( 3\sqrt{1+\frac{2}{3}\omega }+1\right) ^{p_{+}}\left( 3%
\sqrt{1+\frac{2}{3}\omega }-1\right) ^{p_{-}}\left( 9+6\omega \right) ^{-%
\frac{1}{4+3\omega }}t_{m}^{-\frac{2}{4+3\omega }}.  \notag
\end{eqnarray}%
\end{widetext}

The conditions above guarantee the continuity of $a, \Phi, a'$ and $\Phi'$ at any time, including $t_m$. These conditions are not sufficient to guarantee the existence of $a''$ and $\Phi''$ at $t_m$. Since these quantities appear in the EST eq. (\ref{EST cosmo eq1}), it may seem that the latter terms must exist in order for deriving an EST solution. Nevertheless, one can achieve an EST solution even if those second derivatives do not exist, since they only appear in the particular combination  $2a^{\prime\prime }/a+\Phi ^{\prime \prime }/\Phi $. Thence, only the continuity of this quantity, or equivalently of $\left( \Phi a^{2}\right) ^{\prime \prime} $, is required for the solution to make sense. We carefully check these points in Appendix \ref{appB}. Also, one can easily check from (\ref{dust densities}) that the dust density $
\epsilon $ is continuous at $t_{m}$.

With the above, it was possible to derive a nontrivial EST solution. The imposed conditions constrained the model parameters, nonetheless the value of $t_m$ is still arbitrary. This means that there is no way to derive the transition instant $t_m$ from the matching conditions. However, if this transition instant happened in the past of our universe, then a physical imprint of such transition should be left in the cosmological data. That is, from such imprint one can in principle fix the transition instant $t_m$.  Hence, $t_m$ is part of the boundary conditions of the problem.

\bigskip

The case of a perfect fluid having a linear barotropic equation of state is
also tractable and leads to very similar results, as shown in Appendix \ref{appC}.

\bigskip

\begin{figure}[ht]
\centering
\subfloat[Evolution of $a$ and $\Phi$ from GR to ST.]{
    \label{fig:subfig1}
    \includegraphics[width=0.48 \columnwidth]{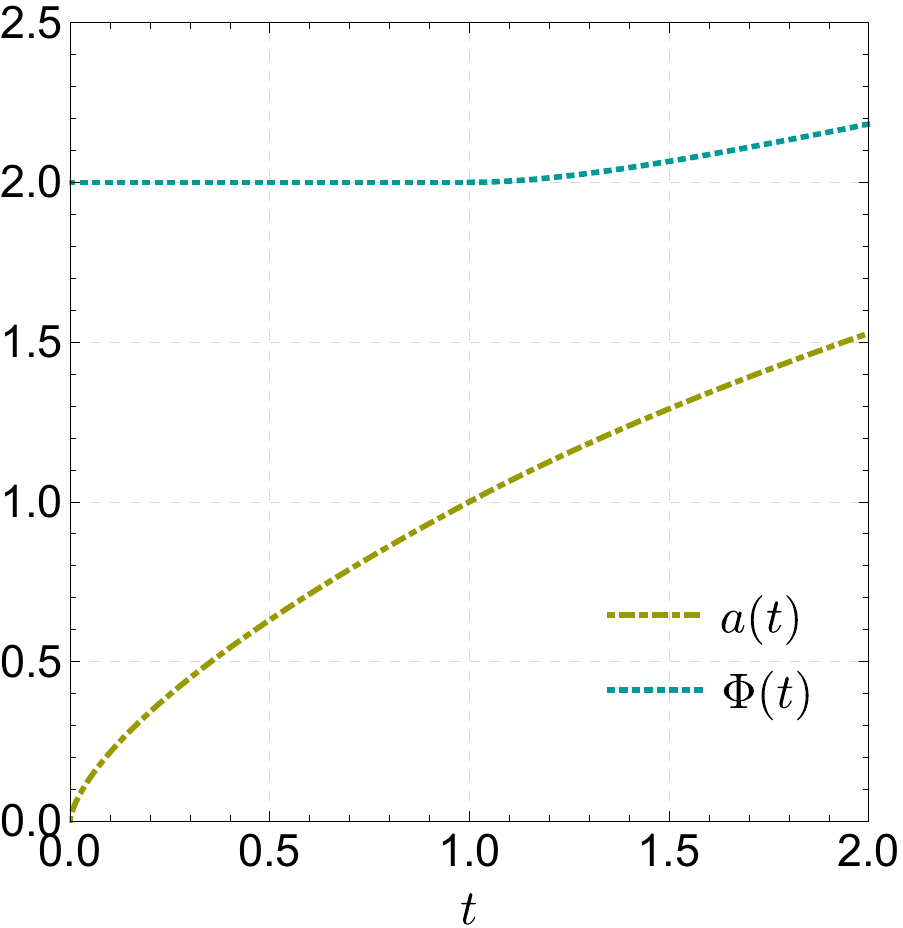}
}
\subfloat[Second derivatives and continuity.]{
    \includegraphics[width=0.48 \columnwidth]{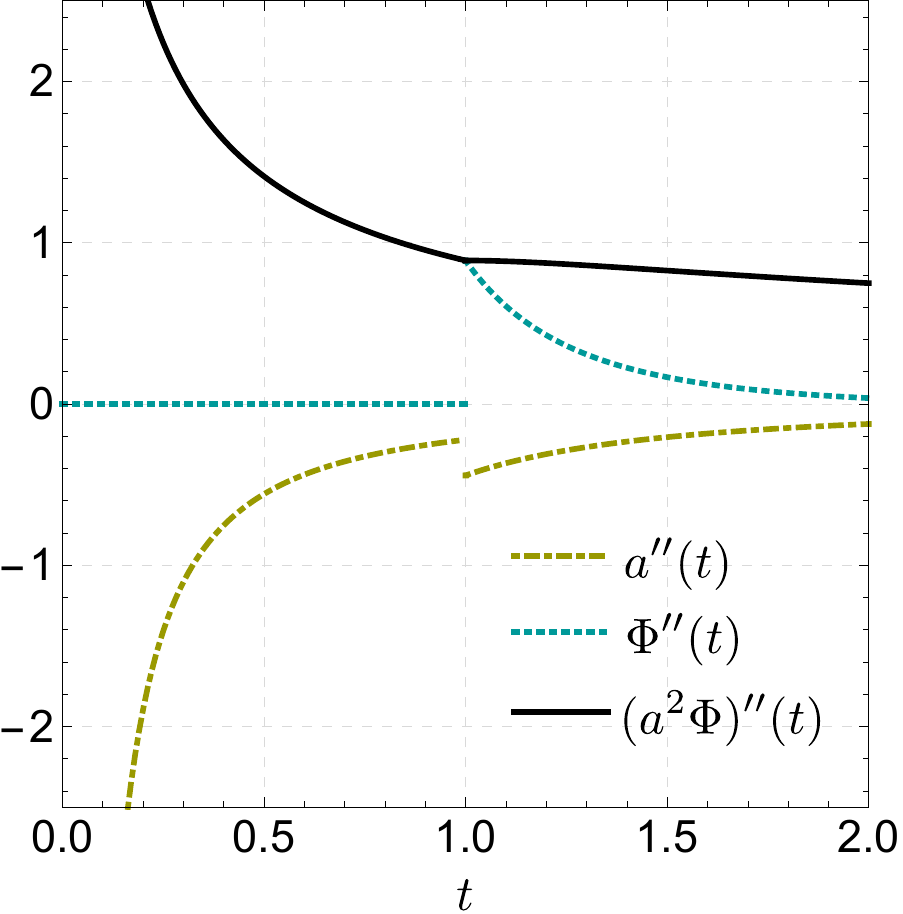}
    \label{fig:subfig2}
}
 \caption{The time evolution of the scale factor $a$ and the scalar field $\Phi$ in the left plot, and the evolution of their second derivatives in the right plot. The transition happens at the instant $t_m=1$, and $a$ was normalized such that $a(t_m)=1$.}
\label{fig1}
\end{figure}

To illustrate these results, let us consider the case $\omega =0$, and
normalise the relevant quantities setting $A=t_{m}=1$ and $C=2$. The BD big
bang time is then $t_{+}=1/4$, and the solution (\ref{GR phase}) and (\ref%
{BD phase}) then reads, using (\ref{q & p with s=-1}), (\ref{BD param tpm})
and (\ref{BD param Bphi})%
\begin{eqnarray}
a_{GR}\left( t\right) =t^{2/3} &  \;\;\;\;\; \text{ and  } \;\;\;\;\;  & \Phi _{GR}\left( t\right)
=2  \label{model numerical}, \\[.12in]
a_{BD}\left( t\right) =\frac{2}{\sqrt{3}}\sqrt{t-\frac{1}{4}} &  \;\;\;\;\;  \text{ and
 } \;\;\;\;\; & \Phi _{BD}\left( t\right) =\frac{2}{\sqrt{3}}\frac{t+\frac{1}{2}}{\sqrt{%
t-\frac{1}{4}}}.  \notag
\end{eqnarray}%
The transition from the GR solution to the ST solution is shown in Figure \ref{fig1}.

This illustrates that, for any $t_{m}$, a solution can be derived by joining the ($t<t_{m}$) GR-like solution to the ($t>t_{m}$) BD-like solution.  It should be noted that the converse is also possible, but it  can only happen at a time where the scalar field experiments an extremum. In this case, the cosmological history begins by an era where a scalar field is acting, resulting in an expansion that occurs faster than the GR case. At $t=t_{m}$, the scalar field is frozen, and the subsequent universe's expansion is governed by GR. Such a possibility may be worth to be considered in the context of inflationary models.

Considering the  solution with $\left( a,\Phi \right) \left(
t<t_{m}\right) =\left( a,\Phi \right) _{GR}$ and $\left( a,\Phi \right)
\left( t>t_{m}\right) =\left( a,\Phi \right) _{BD}$, and using the same
numerical values as in  Figure \ref{fig:subfig1}, the continuity of $\left( \Phi a^{2}\right) ^{\prime \prime }$\ is
 shown in Figure \ref{fig:subfig2}, in accordance with the analytical result presented in Appendix \ref{appB}.

\section{Discussion} \label{sec4}

We showed in this paper that giving an external status to the scalar field
entering a ST theory increases the set of solutions of the theory. Indeed,
both GR and BD solutions are EST solutions, and it is also possible to match pieces of GR
solutions to BD ones in such a way to generate a new class of solutions that
neither belong to GR nor to ST, but satisfies the EST equations. This is
explicitly illustrated by the cosmological example considered in Section
\ref{sec3}. We stress that this new solution is not possible to achieve from pure
BD gravity since the transition from non-GR to GR-like solutions, within BD,
requires $\omega \longrightarrow \infty $, while for an EST theory the
latter can be achieved for arbitrary $\omega $ values.

While GR and ST RW cosmologies constitute well-posed Cauchy problems, this
is false for EST. Indeed, if an EST cosmology is at a GR-like phase at a
certain instant, it is impossible to predict when the change to a BD-like
phase will happen. This means that it is generally necessary to provide more
data than the standard data provided for a Cauchy problem, the transition
time $t_{m}$\ being a part of the boundary conditions to be required.

It is nontrivial that ST gravity with an external scalar field, which we called EST gravity, has solutions that are neither GR nor pure ST, as we have shown here. ST gravities commonly are capable of generating GR solutions under certain limits, while EST gravity has an additional route capable of doing the transition between GR and non-GR local solutions. In this scenario, certain spacetime regions behave as GR solutions, others as ST solutions, while the complete spacetime satisfies the EST equations that are given by eqs. (\ref{Einstein ST}) and (\ref{stress conservation}), or equivalently by eqs. (\ref{Einstein ST}) and (\ref{ext scalar field eq}). This kind of behaviour is akin to screening mechanisms of gravity \cite{v72, kw04, hk10, kmz12}, but the mechanism of EST gravity is of nondynamical nature, depending on the boundary conditions.  We have here explored some cosmological configurations in which nontrivial EST solutions can be derived (Section \ref{sec3}, Appendix \ref{appB}), and we expect to explore non homogeneous and more realistic configurations in a future publication.

In this work, we have specially explored the use of an external scalar as an extension to Brans-Dicke-like actions (\ref{ST action}). In Sec.~\ref{sec3}  explicit results were developed for the particular case of constant $\omega$ and no potential ($\Lambda =0 $).  The extension of this analysis towards nonzero potential is surely important. For example, it would allow to apply the EST proposal to  $f(R)$ theories in such a way that in some spacetime regions either GR or certain $f(R)$ theory would be valid. That is due to the fact that $f(R)$ theories can be expressed by scalar-tensor theories with a non-null potential, and this is valid for diverse $f(R)$ approaches, from the standard one, to Palatini, Hybrid gravity and others \cite{Capozziello:2011et, Capozziello:2012ny, Capozziello:2015lza}. Thus this  EST application would yield an extension of $f(R)$ gravity such that neither GR and nor standard $f(R)$ gravity would be valid in the complete spacetime, but it would be possible to subdivide the spacetime in regions in which either one of the two would be valid. By using the usual correspondence between standard $f(R)$ gravity and ST theories, it is not hard to see that this EST application to $f(R)$ theories yields an $f(R)$ extension in which all the possible metric variations must be considered at the action level, except for those that change the Ricci scalar. Further developments on this EST application to $f(R)$ theories should appear in a future work.

\begin{acknowledgments}
The authors thank Oliver Piattella for stimulating discussions on external
scalar fields and helpful comments. The authors also thank CNPq (Brazil) for
partial financial support. DCR and JCF also thank FAPES (Brazil) for partial
financial support.
\end{acknowledgments}

\appendix

\section{From diffeomorphism invariance to the stress tensor conservation} \label{appA}

Consider the diffeomorphism%
\begin{equation}
x^{\prime a}=x^{a}+\xi ^{a}\left( x^{c}\right) .  \label{diffeo}
\end{equation}%
If the metric's components are considered as functions of the coordinates
(rather than as functions of the point), the difference $\delta g_{ab}\equiv
g_{ab}^{\prime }\left( x\right) -g_{ab}\left( x\right) $ (that is not the
components' variation at a given point) induced by (\ref{diffeo}) reads, at
first order in $\xi $ \cite{mtw73}%
\begin{equation}
\delta g_{ab}=-g_{ac}\partial _{b}\xi ^{c}-g_{bc}\partial _{a}\xi ^{c}-\xi
^{c}\partial _{c}g_{ab}.  \label{metric comp var}
\end{equation}%
Following the same lines, any scalar field $\phi $, while its numerical
value at any point is a gauge invariant, is varied as $\delta \phi \equiv
\phi ^{\prime }\left( x\right) -\phi \left( x\right) =-\xi ^{c}\partial
_{c}\phi $ by (\ref{diffeo}) if $\phi $ is regarded as a function of the
coordinates. The diffeomorphism results in the following matter sector of (%
\ref{ST action}) variation%
\begin{equation}
\delta S_{m}=\dint \left( \frac{\delta \left( \sqrt{-g}L_{m}\right) }{\delta
\Psi }\delta \Psi +\frac{\delta \left( \sqrt{-g}L_{m}\right) }{\delta g_{ab}}%
\delta g_{ab}\right) d^{4}x  \label{matter action var}
\end{equation}%
where the matter Lagrangian $L_{m}$ is defined by $S_{m}=\dint d^{4}x\sqrt{-g%
}L_{m}$, and where the symbol $\frac{\delta l}{\delta F}$\ represents the
usual functional derivative of $l$ w.r.t. the field $F$. If the part $L_{m}$
of the Lagrangian is a scalar, $\dint d^{4}x\sqrt{-g}L_{m}$ is an invariant
integral and $\delta S_{m}$ is then necessarily zero whatever $\xi $. If
none of the matter fields $\Psi $\ is external, each $\Psi $ generates a
Lagrange equation, and since all the $\Psi $ occurrences in the total action (%
\ref{ST action}) enters the matter sector, this Lagrange equation just reads 
$\frac{\delta \left( \sqrt{-g}L_{m}\right) }{\delta \Psi }=0$. Thence, one
gets, using the stress tensor definition $T^{ab}=\frac{2}{\sqrt{-g}}\frac{%
\delta \left( \sqrt{-g}L_{m}\right) }{\delta g_{ab}}$ and (\ref{metric comp
var})%
\begin{equation}
\dint \sqrt{-g}T^{ab}\left( g_{ac}\partial _{b}\xi ^{c}+g_{bc}\partial
_{a}\xi ^{c}+\xi ^{c}\partial _{c}g_{ab}\right) d^{4}x=0.
\label{matter action var2}
\end{equation}%
Since this occurs whatever $\xi $, one gets, after some usual algebra, the
conservation equation (\ref{stress conservation}).

Let us remark that (\ref{stress conservation}) is not recovered if the scalar field $\Phi $ enters $S_{m}$, for two reasons. The first one, relevant in the context of the present paper, is that $\Phi$ is an external field, and then does not generate a Lagrange equation. Second, even in the case were $\Phi $ would be non external, the resulting Lagrange equation $\frac{\delta \left( \sqrt{-g}L\right) }{\delta \Phi }=0$ would not reduce to $\frac{\delta \left( \sqrt{-g}L_{m}\right) }{\delta \Phi }=0$, since the matter sector does not contain all the scalar contributions to the action.

\section{On the continuity of $\left( \Phi a^{2}\right)^{\prime \prime} $} \label{appB}

At $t=t_{m}$, one can easily compute the left and right limits of the second
derivatives, using (\ref{GR phase}) with $t_{0}=0$, (\ref{BD phase}), (\ref%
{q & p with s=-1}), (\ref{BD param tpm}), (\ref{BD param Bphi}) and the
continuity relations (\ref{scalar cont}), (\ref{scalar-der cont}), (\ref%
{scalefact cont}), (\ref{scalefact der cont}). For the scale factor, one
gets%
\begin{eqnarray}
a^{\prime \prime }\left( t_{m}^{-}\right) &=&a_{GR}^{\prime \prime }\left(
t_{m}\right) =-\frac{2A}{9t_{m}^{4/3}},  \\[.1in]
a^{\prime \prime }\left( t_{m}^{+}\right) &=&a_{BD}^{\prime \prime }\left(
t_{m}\right) =-\frac{4\left( 3+\omega \right) A}{9\left( 3+2\omega \right)
t_{m}^{4/3}}, \nonumber
\end{eqnarray}%
and they are different, meaning that $a^{\prime \prime }$ is discontinuous at $%
t_{m}$ (thence that $a^{\prime \prime }\left( t_{m}\right) $ does not
exist). The same occurs for the scalar field%
\begin{eqnarray}
\Phi ^{\prime \prime }\left( t_{m}^{-}\right) &=&\Phi _{GR}^{\prime \prime
}\left( t_{m}\right) =0,  \\[.1in]
\Phi ^{\prime \prime }\left( t_{m}^{+}\right) &=&\Phi _{BD}^{\prime \prime
}\left( t_{m}\right) =\frac{4C}{3\left( 3+2\omega \right) t_{m}^{2}}. \nonumber
\end{eqnarray}%
Now, consider the quantity $\left( \Phi a^{2}\right) ^{\prime \prime }$. In
the GR phase%
\begin{equation}
\left( \Phi a^{2}\right) _{GR}=CA^{2}t^{4/3}.
\end{equation}%
Thence%
\begin{equation}
\left( \Phi a^{2}\right) ^{\prime \prime }\left( t_{m}^{-}\right) =\left(
\Phi a^{2}\right) _{GR}^{\prime \prime }\left( t_{m}\right) =\frac{4CA^{2}}{%
9t_{m}^{2/3}}.
\end{equation}%
In the BD phase%
\begin{equation}
\left( \Phi a^{2}\right) _{BD}=\Phi _{0}B^{2}\left( t-t_{+}\right)
^{2q_{+}+p_{+}}\left( t-t_{-}\right) ^{2q_{-}+p_{-}}.
\end{equation}%
Thence, using (\ref{scalefact der cont}),
\begin{equation}
\left( \frac{\left( \Phi a^{2}\right) _{BD}^{\prime \prime }}{\left( \Phi
a^{2}\right) _{BD}}\right) \left( t_{m}\right) =\frac{16}{9t_{m}^{2}}-\frac{%
2q_{+}+p_{+}}{\left( t_{m}-t_{+}\right) ^{2}}-\frac{2q_{-}+p_{-}}{\left(
t_{m}-t_{-}\right) ^{2}}
\end{equation}%
and one gets%
\begin{equation}
\left( \Phi a^{2}\right) ^{\prime \prime }\left( t_{m}^{+}\right) =\left(
\Phi a^{2}\right) _{BD}^{\prime \prime }\left( t_{m}\right) =\frac{4CA^{2}}{%
9t_{m}^{2/3}}=\left( \Phi a^{2}\right) ^{\prime \prime }\left(
t_{m}^{-}\right) .
\end{equation}%
Since $\left( \Phi a^{2}\right) ^{\prime }\left( t_{m}\right) $\ is
well-defined, $\left( \Phi a^{2}\right) ^{\prime \prime }$ is then
continuous at $t_{m}$, with the value just calculated.

\section{The general linear barotropic case} \label{appC}

In Section \ref{sec3.2} it is presented a cosmological model whose matter content
is a simple dust fluid. This simple kind of matter was used since the main
purpose was to show the existence of nontrivial EST solutions. In this
appendix we show that the existence of nontrivial EST solutions can also be
found in systems with non negligible pressure, in particular we consider a
barotropic fluid. It turns out that the case of a perfect fluid with a
barotropic equation of state having the form $P=n\epsilon $ is also
tractable, leading to similar computations. Indeed, the BD solution reads in
this case \cite{gfr73} (just considering theories with $\omega >-3/2$)%
\begin{eqnarray}
&& a_{BD}\left( t\right) =\widetilde{B}^{\frac{1}{1-n}}\left( \widetilde{t}-%
\widetilde{t}_{+}\right) ^{Q_{+}}\left( \widetilde{t}-\widetilde{t}%
_{-}\right) ^{Q_{-}}  \label{BD barotrop sol}, \\[.1in]
& & \Phi _{BD}\left( t\right) =\widetilde{\Phi }_{0}\left( \widetilde{t}-%
\widetilde{t}_{+}\right) ^{P_{+}}\left( \widetilde{t}-\widetilde{t}%
_{-}\right) ^{P_{-}},  \\[.1in]
&& \epsilon _{BD}\left( t\right) a_{BD}\left( t\right) ^{3+3n} =\widetilde{M}
\end{eqnarray}%
with%
\begin{eqnarray}
Q_{\pm } &=&\frac{1+\left( 1-n\right) \omega \pm \sqrt{1+\frac{2}{3}\omega }%
}{4-6n+3\left( 1-n\right) ^{2}\omega }  \label{Q barotrop}, \\[.1in]
P_{\pm } &=&\frac{1-3n\mp 3\left( 1-n\right) \sqrt{1+\frac{2}{3}\omega }}{%
4-6n+3\left( 1-n\right) ^{2}\omega }, 
\end{eqnarray}%
where $\widetilde{t}$ is a conformal time defined by%
\begin{equation}
dt=a^{3n}d\widetilde{t}.  \label{conformal time}
\end{equation}%
The constants $\widetilde{B}$, $\widetilde{\Phi }_{0}$, $\widetilde{t}_{-}$, $\widetilde{t}%
_{+}$ and $\widetilde{M}$\ are the five integration constants. The GR
solution $a_{GR}\left( t\right) \propto \left( t-t_{0}\right) ^{2/\left(
3+3n\right) }$ can be written, using the same conformal time, and choosing
the origin of time such that $t_{0}=0$,
\begin{eqnarray}
a_{GR}\left( t\right) &=&\widetilde{A}\widetilde{t}^{2/\left( 3-3n\right) }
\label{GR barotrop sol}, \\[.1in]
\Phi _{GR}\left( t\right) &=&\widetilde{C}, \\[.1in]
\epsilon _{GR}\left( t\right) a_{GR}\left( t\right) ^{3+3n} &=&\widetilde{M}.
\end{eqnarray}%
The continuity of the scale factor and the scalar and of their derivatives
with respect to $t$ result in the continuity of their derivatives with
respect to the conformal time $\widetilde{t}$, since $\frac{d}{d\widetilde{t}%
}=\frac{dt}{d\widetilde{t}}\frac{d}{dt}=a^{3n}\frac{d}{dt}$. These
continuities at $\widetilde{t}_{m}$ lead to a system that can be put on a
form very close to (\ref{scalar cont})-(\ref{scalefact der cont}),
\begin{equation}
\widetilde{\Phi }_{0}\left( \widetilde{t}_{m}-\widetilde{t}_{+}\right)
^{P_{+}}\left( \widetilde{t}_{m}-\widetilde{t}_{-}\right) ^{P_{-}}=%
\widetilde{C},  \label{barotrop scalar cont}
\end{equation}%

\begin{equation}
\frac{P_{+}}{\widetilde{t}_{m}-\widetilde{t}_{+}}+\frac{P_{-}}{\widetilde{t}%
_{m}-\widetilde{t}_{-}}=0,  \label{barotrop scalar-der cont}
\end{equation}%

\begin{equation}
\widetilde{B}\left( \widetilde{t}_{m}-\widetilde{t}_{+}\right) ^{\left(
1-n\right) Q_{+}}\left( \widetilde{t}_{m}-\widetilde{t}_{-}\right) ^{\left(
1-n\right) Q_{-}}=\widetilde{A}\widetilde{t}_{m}^{2/3},
\label{barotrop scalefact cont}
\end{equation}%

\begin{equation}
\frac{\left( 1-n\right) Q_{+}}{\widetilde{t}_{m}-\widetilde{t}_{+}}+\frac{%
\left( 1-n\right) Q_{-}}{\widetilde{t}_{m}-\widetilde{t}_{-}}=\frac{2}{3%
\widetilde{t}_{m}}.  \label{barotrop scalefact-der cont}
\end{equation}%
The solution for the BD constants is then the same as for the dust case,
just replacing the exponents $p_{\pm }$ and $q_{\pm }$\ by $P_{\pm }$ and $%
\left( 1-n\right) Q_{\pm }$ respectively.

\section{A mechanical example with an external coordinate} \label{appMec}
This appendix aims to clarify the physical meaning of external fields by providing an additional example within a very simple context, that of a planar double pendulum. 

Before considering the case with external coordinates, consider the usual Lagrangian for the double pendulum subject to a uniform gravitational acceleration whose norm is denoted by $g$ \cite{0750628960},
\begin{equation}
	\label{eq:LO}
	L(\theta_i,\dot \theta_i)=\frac{1}{2} \dot{\theta }_1^2 l_1^2 \left(m_1+m_2\right)+\frac{1}{2} \dot{\theta }_2^2 l_2^2 m_2+\dot{\theta }_1 \dot{\theta }_2 l_1 l_2 m_2 \cos \left(\theta _1-\theta _2\right)  - V(\theta_i),
\end{equation}
with
\begin{equation}
	V(\theta_i) = - g l_2 m_2 \cos \left(\theta _2\right) -g l_1 \left(m_1+m_2\right) \cos \left(\theta _1\right),
\end{equation}
where $\theta_i$ is used as a shorthand notation for $\theta_1, \theta_2$, the particle masses are denoted by $m_1$ and $m_2$, $m_1$ is supported by a wire of negligible mass of length $l_1$, $m_1$ and $m_2$ are connected by another wire of length $l_2$, $\theta_1$ and $\theta_2$ are the position angles of each of the particles in regard to a straight line parallel to the gravitational acceleration vector $\boldsymbol{g}$.

The equations of motion read 
\begin{eqnarray}
&& \left (1 + \frac{m_2}{m_1} \right)\frac{l_1}{l_2} \ddot{\theta }_1 + \frac{m_2}{m_1} \ddot{\theta }_2 \cos \left(\theta _1-\theta _2\right) + \frac{m_2}{m_1}\dot{\theta }_2^2  \sin \left(\theta _1-\theta _2\right)+ \frac{g}{l_2} \left(1 + \frac{m_2}{m_1}\right) \sin \left(\theta _1\right) = 0, \\[.2in] 
&& \ddot{\theta }_2 \frac{l_2}{l_1}+ \ddot{\theta }_1 \cos \left(\theta _1-\theta _2\right)- \dot{\theta }_1^2 \sin \left(\theta _1-\theta _2\right) +\frac{g}{l_1} \sin \left(\theta _2\right) = 0.
\end{eqnarray}
Considering that $m_1$ is sufficiently large, the above equations can be written as
\begin{eqnarray}
\label{eq:theta1}
&& l_1\ddot{\theta }_1=-g \sin \left(\theta_1\right), \\[.2in] 
\label{eq:theta2}
&& \frac{l_2}{l_1}\ddot{\theta }_2 = -\frac{g \sin \left(\theta _2\right)}{l_1} - \ddot{\theta }_1 \cos \left(\theta _1-\theta _2\right)+ \dot{\theta }_1^2 \sin \left(\theta _1-\theta _2\right).
\end{eqnarray}
In this limit, the second pendulum has no effect on the first one. Alternatively, one can also consider the regime $m_1 \gg m_2$ right in the Lagrangian (or the action), and the resulting Lagrangian reads
\begin{eqnarray}
	L (\theta_i, \dot \theta_i) &=& m_2 \left\{ \frac{1}{2} \dot{\theta }_2^2 l_2^2+\dot{\theta }_1 \dot{\theta }_2 l_1l_2 \cos \left(\theta _1-\theta _2\right) + g l_2 \cos(\theta_2)\right\} +\nonumber \\[.2in] 
	&& +\left(m_1+m_2\right) \left(g l_1 \cos \left(\theta _1\right)+\frac{1}{2} \dot{\theta }_1^2 l_1^2\right) \nonumber \\[.2in]
	&= & m_2 L_2(\theta_i,\dot \theta_i) + (m_1 + m_2) L_1(\theta_1,\dot \theta_1) \nonumber \\[.2in]
	&\approx & m_2 L_2(\theta_2,\dot \theta_2, t) + m_1 L_1(\theta_1,\dot \theta_1).
\end{eqnarray}
The last approximation is constituted by two parts, the first is a trivial one in which it was used that $m_1 + m_2 \approx m_1$. The second part refers to the replacement $L_2(\theta_i, \dot \theta_i) \rightarrow L_2(\theta_2, \dot \theta_2, t)$, in which $L_2$ becomes a function of the second particle only, and the occurrence of $\theta_1$ inside $L_2$ is seen just as function of the time variable $t$, not a generalised coordinate with the same status of $\theta_2$. This type of decoupling on the dynamics of the particles 1 and 2 is expected since, by requiring that $m_1$ is sufficiently large, the extremum of the total action $\int L dt$ should be achieved only when both $m_2 \int L_2 dt$ and $m_1\int L_1 dt$ are independently extremized. Hence, a consistent approach is the following: from the Euler-Lagrange equations of $L_1$ one derives $\theta_1(t)$; and with this result, the $\theta_1$ solution is inserted into $L_2$, which is then used to derive the $\theta_2$ solution.

Another way of justifying  why it is correct and necessary to use $L_2(\theta_i, \dot \theta_i) \rightarrow L_2(\theta_2, \dot \theta_2, t)$, when the regime $m_1 \gg m_2$ is considered, comes from the fact that the Euler-Lagrange equations of $L_1(\theta_1, \dot \theta_1)$ leads to the eq. (\ref{eq:theta1}), while the Euler-Lagrange equation of $L_2(\theta_2,\dot \theta_2, t)$ leads to eq. (\ref{eq:theta2}). Moreover, if one considers $L_2$ as a function of $\theta_1$ and derive its corresponding Euler-Lagrange equation, the derived equation is not compatible with eq. (\ref{eq:theta2}). Hence it is wrong to deal with $\theta_1$ in $L_2$ as if it were a standard coordinate: $\theta_1$ is an external coordinate in $L_2(\theta_2, \dot \theta_2,t)$.

\bigskip
\noindent
In the following a generalisation  of the picture above presented will be considered. Eventually, the dynamics of the external coordinate may not be well known, that is, the Lagrangian $L_1$ may be more complex. Since eq. (\ref{eq:theta2}) does not depend on $L_1$, one can use $L_2$ in a more general situation in which $L_1$ is unknown. In this setting, it will not be possible in general to derive the full solution of both $\theta_1$ and $\theta_2$, unless additional information is provided. Nonetheless we stress that it is correct to use eq. (\ref{eq:theta2}) even though $L_1$, and consequently $\theta_1(t)$, is unknown. 

	Consider the Lagrangian
\begin{equation}
	\tilde L(\theta_i, \dot \theta_i)  = L(\theta_i, \dot \theta_i) + k L_3( \theta_1, \dot \theta_1, \ddot \theta_1, ...),
\end{equation}
where $L$ is given by eq. (\ref{eq:LO}), $k$ is a dimensionless quantity, and $L_3$ is any function that depends only on $\theta_1$ and its derivatives (it may or may not depend on higher derivatives). Here it is not assumed that $m_1 \gg m_2$. Nonetheless, for $k$ sufficiently large the same argument used in the case in which $m_1$ is sufficiently large can be used. Therefore the extremum of the action $\int \tilde L dt$ will be achieved only when both $\int L dt$ and $\int L_3 dt$ are  extremized, thus, in the regime $k \gg 1$, one can write
\begin{equation}
	\tilde L(\theta_i, \dot \theta_i) \approx L(\theta_2, \dot \theta_2, t) + k L_3( \theta_1, \dot \theta_1, \ddot \theta_1, ...),
\end{equation}
where the explicit time dependence of $L$ comes from the solution of $\theta_1(t)$, which in turn can be directly derived from the equations of motion of $\int L_3 dt$. It should be noted that even if the Lagrangian $L_3$ is unknown, one can use the equation of motion of $L_2$ with respect to $\theta_2$, while $\theta_1$ is seen as an external coordinate. In this case, $\theta_1(t)$ will be an arbitrary function.

\section{A first test of an EST model with SN data} \label{appE}

The main intention of the present paper is to show the existence of nontrivial solutions within the EST approach.   It has been shown in other sections that the matching between GR and BD solutions, in the context of the EST approach, is possible. Although, for the moment, the resulting cosmological model is too simplified to constitute a realistic scenario, it is already possible to test the configuration here obtained with SN data. This is done here both to show that it is possible and as a first step towards testing EST models. To this end, we use the binned JLA sample \cite{jla}. The advantage of such sample is related to the binned process which lead to less important effects due to different calibration methods.

We match the flat CDM model of GR with the flat BD solutions used above. Of course, the observational test disfavour strongly the CDM with respect to the $\Lambda$CDM model. We use the expressions derived above, and consider three free parameters, the matching time $t_m$, the parameter $\omega$ and the Hubble parameter $h$. As usual, the fit is done looking for the values of these parameters that lead to the global minimum of $\chi^2$. 

\begin{figure}[h]
\centering
\includegraphics[width=9cm]{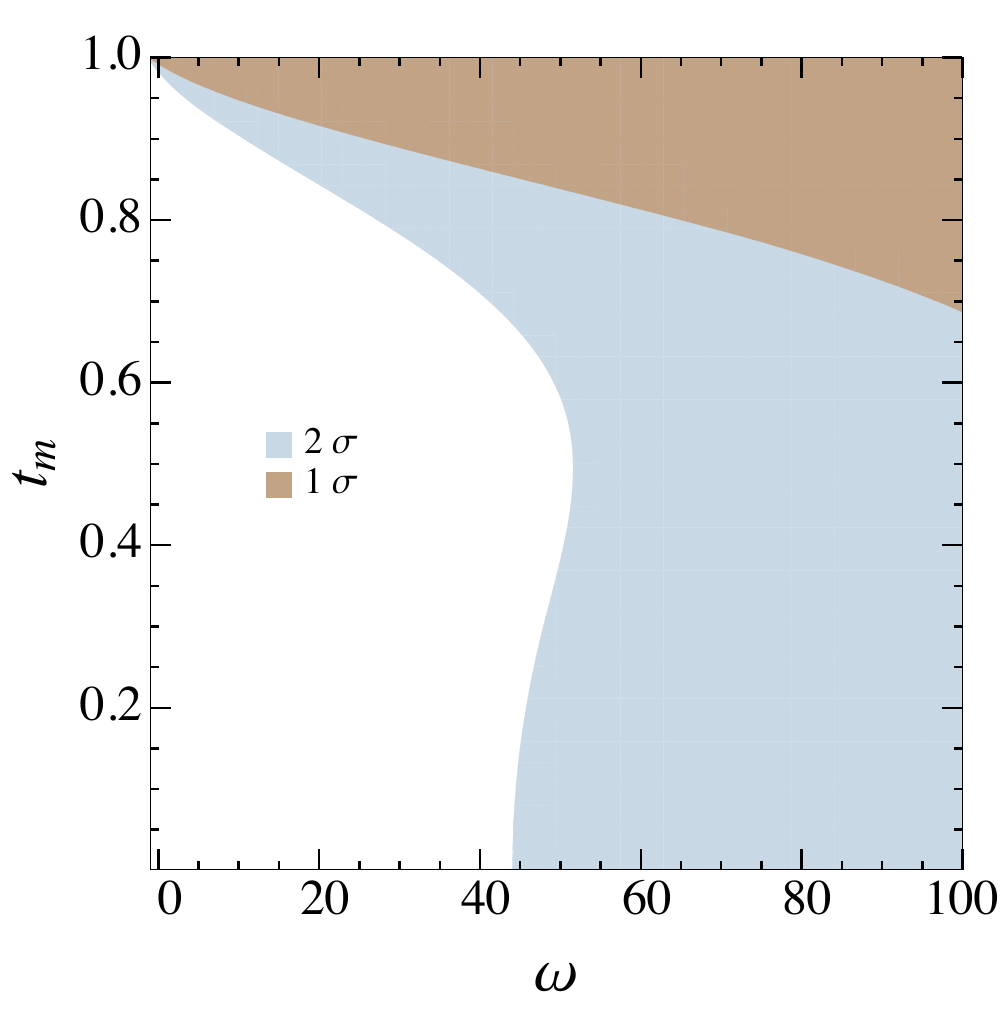}
\caption{The $1\sigma$ and $2 \sigma$ confidence regions in the $(\omega, t_m)$ space. The SN fit was done with three free parameters, the latter two and the Hubble parameter ($h$). For the result above, $h$ was marginalised.} \label{fig:f1pb}
\end{figure}

The results indicate that the global maximum of probability (or global minimum of $\chi^2$) occurs for $t_m = t_0$, where $t_0$ is the present time, implying that the data favours no phase transition and the universe is still in the GR phase. The derived value of the minimum $\chi^2$ is not small in comparison to the standard cosmological model $\Lambda$CDM, but this is expected since this model includes no dark energy. The resulting fit is just as good as that of a pure CDM model. Likewise pure CDM, the minimum $\chi^2$, for this set of data,  is $\chi^2 \approx 350$.  There is a local maximum for $t_m = 0$ corresponding to a pure BD phase, for the latter, the corresponding value is $\chi^2 \approx 353$. For comparison purposes, the minimum $\chi^2$ for $\Lambda$CDM is 29. The $1 \sigma$ and $2 \sigma$ confidence regions  are displayed in fig. (\ref{fig:f1pb}). As expected, the higher is $\omega$, more similar the dynamics of the BD and the GR phases become, and hence it becomes more probable for the transition to  happen in an instant before $t_0$. 

A complete statistical study would imply to consider other tests and to introduce a mechanism to accelerate the expansion of the universe. This has in part been done in Ref. \cite{brasil} in the context of the usual BD theory by including the possibility of $\omega < - 4/3$, what has been excluded in our analysis. In that reference, it has been found, comparing the pure BD flat matter dominated phase with the $\Lambda$CDM, that the BD model is slightly favoured for $\omega \lesssim - 4/3$.

\begin{thebibliography}{}

\bibitem{cf11} S.\ Capozziello, V. Faraoni, \textit{Beyond Einstein's gravity} (Fundamental Theories of Physics, volume 170, Springer,
2011)

\bibitem{f04} V.\ Faraoni, \textit{Cosmology in scalar-tensor
gravity} (Fundamental theories of Physics, volume 139, Kluwer Academic
Publishers, 2004)


\bibitem{fm03} Y. Fujii, K.\ Maeda, \textit{The scalar-tensor theory of gravitation} (Cambridge University Press, 2003).

\bibitem{w14} C.\ M.\ Will, \textit{The confrontation between general relativity and experiments, }in www.livingreviews.org/Irr-2014-4
(living reviews in relativity)

\bibitem{hk10} K.\ Hinterbichler, J.\ Khoury, Phys. Rev. Lett. \textbf{104}, 231301 (2010)

\bibitem{kmz12} T. S. Koivisto, D.\ F.\ Mota, M.\ Zumalacarregui, Phys. Rev. Lett. \textbf{109}, 241102 (2012)

\bibitem{kw04} J.\ Khoury, A.\ Weltman, Phys. Rev. Lett. \textbf{93}, 171104 (2004)

\bibitem{v72} A.\ Vainshtein, Phys. Lett. B \textbf{39}, 393 (1972)

\bibitem{rw04} M.\ Reuter, H.\ Weyer, Phys. Rev. D \textbf{69}, 104022 (2004); Phys. Rev. D \textbf{70}, 124028 (2004)

\bibitem{rls10} D.C.\ Rodrigues, P.\ S.\ Letelier, I.\ L.\ Shapiro, JCAP \textbf{04}, 020 (2010)

\bibitem{sss05} I.\ L.\ Shapiro, J.\ Sol\`{a}, H.\ Stefancic, JCAP \textbf{01}, 012 (2005)

\bibitem{rcp15}  D.~C.~Rodrigues, B.~Chauvineau and O.~F.~Piattella,
  JCAP {\bf 1509}, no. 09, 009 (2015)
  [arXiv:1504.05119 [gr-qc]].

\bibitem{Hassan:2011tf} 
  S.~F.~Hassan, R.~A.~Rosen and A.~Schmidt-May,
  JHEP {\bf 1202}, 026 (2012)
  [arXiv:1109.3230 [hep-th]].


\bibitem{Isham:1971gm} 
  C.~J.~Isham, A.~Salam and J.~A.~Strathdee,
  Phys.\ Rev.\ D {\bf 3}, 867 (1971).
  

\bibitem{Damour:2002wu} 
  T.~Damour, I.~I.~Kogan and A.~Papazoglou,
  Phys.\ Rev.\ D {\bf 66}, 104025 (2002)
  [hep-th/0206044].

  

\bibitem{deRham:2010kj} 
  C.~de Rham, G.~Gabadadze and A.~J.~Tolley,
  Phys.\ Rev.\ Lett.\  {\bf 106}, 231101 (2011)
  [arXiv:1011.1232 [hep-th]].  

  \bibitem{Hassan:2011vm} 
  S.~F.~Hassan and R.~A.~Rosen,
  JHEP {\bf 1107}, 009 (2011)
  [arXiv:1103.6055 [hep-th]].
  
  
  \bibitem{Weinberg:1988cp} 
  S.~Weinberg,
  Rev.\ Mod.\ Phys.\  {\bf 61}, 1 (1989).

  \bibitem{Unruh:1988in} 
  W.~G.~Unruh,
  Phys.\ Rev.\ D {\bf 40}, 1048 (1989).

 \bibitem{Henneaux:1989zc} 
  M.~Henneaux and C.~Teitelboim,
  Phys.\ Lett.\ B {\bf 222}, 195 (1989).
 
  
  \bibitem{Ellis:2010uc} 
  G.~F.~R.~Ellis, H.~van Elst, J.~Murugan and J.~P.~Uzan,
  Class.\ Quant.\ Grav.\  {\bf 28}, 225007 (2011)
  [arXiv:1008.1196 [gr-qc]].
  
  
  \bibitem{Kluson:2014esa} 
  J.~Kluson,
  Phys.\ Rev.\ D {\bf 91}, no. 6, 064058 (2015)
  [arXiv:1409.8014 [hep-th]].
  
 
  \bibitem{FernandezCristobal:2014jca} 
  J.~M.~Fern\'andez Crist\'obal,
  Annals Phys.\  {\bf 350}, 441 (2014).


\bibitem{bt13} C.\ G. B\"{o}hmer, N. Tamanini, Foundations of Physics\ \textbf{43}, 1478 (2013)

\bibitem{w93} C. M. Will, \textit{Theory and experiment in gravitational physics} (Cambridge University Press, 1993)


\bibitem{bd61} C. Brans, R. H. Dicke, Phys. Rev. \textbf{124}, 925 (1961)

\bibitem{gfr73} L. E. Gurevich, A. M. Finkelstein, V. A. Ruban, Astrophys. and Sp. Sc.\ \textbf{22}, 231 (1973)


\bibitem{c07} B. Chauvineau, Gen. Relativ. Gravit.\ \textbf{39}, 297 (2007)

\bibitem{c03} B. Chauvineau, Class. Quantum Grav. \textbf{20}, 2617 (2003)

\bibitem{bv81} W. B. Bonnor, P. A. Vickers, Gen. Relativ. Gravit.\ \textbf{13}, 29 (1981)

\bibitem{mtw73} C. W. Misner, K. S. Thorne, J. A. Wheeler, \textit{Gravitation} (Freeman, San Francisco, 1973)


\bibitem{h85} F. Hammer, Astron. Astrophys. \textbf{152}, 262 (1985)


\bibitem{nh84} L.\ Nottale, F. Hammer, Astron. Astrophys. \textbf{141}, 144 (1984)

\bibitem{b00} W. B. Bonnor, Class. Quantum Grav. \textbf{17}, 2739 (2000)

\bibitem{nkty14} S.\ Nishi, T.\ Kobayashi, N.\ Tanahashi, M. Yamaguchi, JCAP \textbf{03}, 008 (2014)

\bibitem{0750628960} L. D. Landau,  E. M. Lifshitz, {\it Mechanics, Third Edition: Volume 1 (Course of Theoretical Physics S)}, Butterworth-Heinemann (1976)

\bibitem{jla} M. Betoule et al. [SDSS Collaboration], Astron. Astrophys. {\b 568}, A22 (2014) [arXiv:1401.4064
[astro-ph.CO]].

\bibitem{brasil} J.C. Fabris, S.V.B. Goncalves and R. de Sa Ribeiro, Grav.\&Cosmol. {\bf 12},49(2006).

\bibitem{Capozziello:2011et} S.~Capozziello and M.~De Laurentis,
  Phys.\ Rept.\  {\bf 509}, 167 (2011)
  doi:10.1016/j.physrep.2011.09.003
  [arXiv:1108.6266 [gr-qc]].
  
\bibitem{Capozziello:2012ny} 
  S.~Capozziello, T.~Harko, T.~S.~Koivisto, F.~S.~N.~Lobo and G.~J.~Olmo,
  JCAP {\bf 1304}, 011 (2013)
  doi:10.1088/1475-7516/2013/04/011
  
  \bibitem{Capozziello:2015lza} 
  S.~Capozziello, T.~Harko, T.~S.~Koivisto, F.~S.~N.~Lobo and G.~J.~Olmo,
  Universe {\bf 1}, no. 2, 199 (2015)
  doi:10.3390/universe1020199
  [arXiv:1508.04641 [gr-qc]].


\end{thebibliography}
\end{document}